\documentclass[prx,amssymb,twocolumn,tightenlines,showpacs]{revtex4-1}

\usepackage{graphicx}
\usepackage{amsmath}
\usepackage{verbatim}
\usepackage{subfigure}

\begin{document}

\title{Entanglement Dynamics in the Presence of Controlled Unital Noise}

\author{A. Shaham}
\affiliation{Racah Institute of Physics, Hebrew University of
Jerusalem, Jerusalem 91904, Israel}
\author{A. Halevy}
\affiliation{Racah Institute of Physics, Hebrew University of
Jerusalem, Jerusalem 91904, Israel}
\author{L. Dovrat}
\affiliation{Racah Institute of Physics, Hebrew University of
Jerusalem, Jerusalem 91904, Israel}
\author{E. Megidish}
\affiliation{Racah Institute of Physics, Hebrew University of
Jerusalem, Jerusalem 91904, Israel}
\author{H. S. Eisenberg}
\affiliation{Racah Institute of Physics, Hebrew University of
Jerusalem, Jerusalem 91904, Israel}


\begin{abstract}
Quantum entanglement is notorious for being a very fragile resource.
Significant efforts have been put into the study of entanglement
degradation in the presence of a realistic noisy environment. Here,
we present a theoretical and an experimental study of the
decoherence properties of entangled pairs of qubits. The
entanglement dynamics of maximally entangled qubit pairs is shown to
be related in a simple way to the noise representation in the Bloch
sphere picture. We derive the entanglement level in the case when
both qubits are transmitted through any arbitrary unital Pauli
channel, and compared it to the case when the channel is applied
only to one of the qubits. The dynamics of both cases was verified
experimentally using an all-optical setup. We further investigated
the evolution of partially entangled initial states. Different
dynamics was observed for initial mixed and pure states of the same
entanglement level.
\end{abstract}

\maketitle

\section{Introduction}
Quantum entanglement is an essential ingredient in many quantum
information tasks. It is vital for the implementation of quantum
protocols such as quantum teleportation and dense coding, as well as
for other computational schemes \cite{Nielsen}. Decoherence -- the
undesired coupling of a quantum system to other non-accessible
systems, results in quantum noise which reduces the degree of
entanglement that the system of interest possesses. This in turn can
hinder the success of quantum information protocols. Therefore, the
characterization of entanglement dynamics under the influence of
decohering processes is required for any future realization of these
quantum information applications.

There are two main approaches for studying the dynamics of
entanglement. In the first, a specific model which is a result of a
specific physical implementation of noise is considered
\cite{Zyczkowski_2001,Yu_2004,Dodd_2004,Cabello_2005}. In the second
approach, dynamics is studied more generally, based on the knowledge
of the noise parameters and the initial state
\cite{Puentes_2007,Konrad_2008,Jimenez_2009,Xu_2009}. Some previous
works have focused on the specific moment where entanglement
disappears, referring to it as the sudden death point if
entanglement vanishes while local coherence still prevails
\cite{Yu_2006, Almeida_ESD}. Other works addressed a more general
binary question: does a given noisy channel preserve or break
entanglement? \cite{Ruskai,Ziman_2005,Filippov}. Among the above
works, a few have dealt with the case where the noise is unital,
i.e., mapping the maximally mixed state onto itself.
\cite{Zyczkowski_2001,Ruskai,Ziman_2005,Filippov}.

In this work, we investigate the entanglement dynamics of qubit
pairs transmitted through controlled uncorrelated and local unital
noisy channels. The dynamics of two maximally entangled qubits that
are both affected by the channel is derived. We show that the
entanglement level of the system, quantified by the concurrence
measure, is described by simple relations of the important
parameters of the noise. These relations are verified experimentally
using an all-optical setup. Furthermore, by generating different
initial states, it is shown how entanglement evolution is affected
by the amount of entanglement in the initial state, and its
mixedness.

\section{Theoretical model}
The entanglement of a qubit pair is commonly quantified using the
concurrence measure \cite{Wootters} defined as
$C(\hat{\rho})=\textrm{max}\{0,Q(\hat{\rho})\}$, where
$Q=\sqrt{\lambda_0}-\sqrt{\lambda_1}-\sqrt{\lambda_2}-\sqrt{\lambda_3}$,
$\lambda_i$ are the ordered positive eigenvalues of
$\Lambda=\hat{\rho}(\sigma_y\otimes\sigma_y)\hat{\rho}^*(\sigma_y\otimes\sigma_y)$,
$\hat{\rho}^*$ is the complex conjugate of the density matrix
$\hat{\rho}$ of the state in the computational basis and $\sigma_y$
is the second Pauli matrix. The concurrence is one for maximally
entangled states such as the Bell states
$|\psi^\pm\rangle=(|01\rangle\pm|10\rangle)/\sqrt{2}$ and
$|\phi^\pm\rangle=(|00\rangle\pm|11\rangle)/\sqrt{2}$, and zero for
separable states.

Consider a quantum channel that acts on a single-qubit state
$\hat{\rho}$. The operation of the channel can be uniquely
described by a completely positive map $\mathcal{E}$, using the
elements of the process matrix $\chi$
\begin{equation}\label{process_definition}
\mathcal{E}(\hat{\rho})=\sum_{m,n}\chi_{mn}\hat{E}_{m}\hat{\rho}\hat{E}_{n}^\dag,
\end{equation}
where $\hat{E}_m$ are called Kraus operators and span the vector
space of $\hat{\rho}$. $\chi$ is positive, Hermitian, and satisfies
$\textrm{Tr}(\chi)=1$ (i.e., the channel is nondissipative). In the
case of a quantum system in $n$-dimensional Hilbert space, $\chi$ is
a $n^2\times n^2$ matrix. In addition, the channel can be
geometrically represented as the mapping of the surface of the Bloch
sphere onto a smaller contained ellipsoid surface \cite{Nielsen}.

If the channel is unital (i.e., $\mathcal{E}(\hat{I})=\hat{I}$), the
sphere surface and the mapped ellipsoid are concentric. As was shown
in Ref. \citenum{King_2001}, the mapping operation of unital
channels can be understood as the implementation of two unitary
rotations $\{U,V\}$, along with a three-parameter simpler map
$\mathcal{E}_D$:
\begin{equation}\label{process_as_rotations_and_shrinking}
\mathcal{E}(\hat{\rho})=U\mathcal{E}_D(V\hat{\rho}V^\dagger)U^\dagger.
\end{equation}
The map $\mathcal{E}_D$ can be described using three parameters
$\{R_1, R_2, R_3\}$, which are the lengths of the primary axes of
the mapped ellipsoid. For unital channels, the process matrix that
describes the $\mathcal{E}_D$ operation is $\chi_D$, the
diagonalization of the process matrix $\chi$. The radii $\{R_1, R_2,
R_3\}$ are related to the eigenvalues of $\chi$ by
\begin{equation}\label{Radii chi eigenvalues}
R_i=\chi_0+\chi_i-\chi_j-\chi_k\,,
\end{equation}
where $i\neq{j}\neq{k}\neq0$ \cite{King_2001}. A negative value of
$R_i$ is interpreted as an inversion of the mapped ellipsoid with
respect to a plane perpendicular to $R_i$. The complete positivity
of $\chi$ is equivalent to the requirements that
$|R_i\pm{R_j}|\leq|1\pm{R_k}|$ \cite{King_2001}. These inequalities
define an allowed tetrahedral volume within the three dimensional
radii space \cite{King_2001}.

A unital channel that does not include rotations (i.e., $U=V=I$) is
called a Pauli channel. When the $\hat{E}$ matrices are the
$\sigma_0$ identity matrix and the $\sigma_1$, $\sigma_2$, and
$\sigma_3$ Pauli matrices, the $\chi$ matrix that describes the
Pauli channel is the diagonal $\chi_D$ matrix. Denoting the
correseponding eigenvalues of $\chi_D$ by $\{\chi_0,..,\chi_3\}$, we
write the probability for a change in the initial state as
\begin{equation}\label{change_probability_of_a_process}
P=\sum_{i=1}^{3}\chi_{i}=1-\chi_0\,.
\end{equation}

We analyze the case when a qubit pair is initially prepared in a
maximally entangled Bell state $|\psi_B\rangle$. A unital Pauli
channel, which is designated by the symbol \$ and represented by a
diagonal matrix $\chi_D$, is then applied with a probability $P$ to
one of the two qubits (i.e., $\mathcal{E}=(I\otimes\$$)).
Calculating the output state
$\mathcal{E}(|\psi_B\rangle\langle\psi_B|)$ using Eq.
(\ref{process_definition}), we obtain that the eigenvalues of the
corresponding $\Lambda[(I\otimes\$)|\psi_B\rangle\langle\psi_B|]$
matrix are $\{\chi_0^2,\chi_1^2,\chi_2^2,\chi_3^2\}$. If
$\lambda_0=\chi_0^2$ is the maximal eigenvalue, the concurrence in
terms of $P$ is written using Eq.
\ref{change_probability_of_a_process} as:
$C=\textrm{max}\{\chi_0-\chi_1-\chi_2-\chi_3,0\}=\textrm{max}\{1-2P,0\}$.

Rewriting the concurrence as a function of the primary radii $R_i$,
we obtain: $C=\textrm{max}\{(R_1+ R_2+R_3-1)/2,0\}$. Generalizing
this relation to the case where the maximal eigenvalue of $\Lambda$
is not $\lambda_0$, the concurrence becomes
\begin{equation}\label{one_qubit_dynamics}
C[\mathcal{E}(|\psi_B\rangle\langle\psi_B|)]=\textrm{max}\{(|R_1|+
|R_2|+|R_3|-1)/2,0\}.
\end{equation}
This equation describes the concurrence evolution when the noise is
unital \cite{Ziman_2005}. An immediate result from Eq.
(\ref{one_qubit_dynamics}) is the entanglement breaking condition
\cite{Ruskai}: entanglement disappears when the channel satisfies
$|R_1|+|R_2|+|R_3|\leq1$. As unitary local rotations do not change
the amount of entanglement, Equation (\ref{one_qubit_dynamics}) is
valid not just for the Bell states, but also for any maximally
entangled initial state.

A second situation that we would like to study is when the same
local unital Pauli process $\$$ is applied to both qubits of
$|\psi_B\rangle$ (i.e., $\mathcal{E}=(\$\otimes\$$)). Calculating
the concurrence of the output state as a function of the eigenvalues
of $\chi$ $\{\chi_0,..,\chi_3\}$, we find that the eigenvalues of
$\Lambda[(\$\otimes\$)|\psi_B\rangle\langle\psi_B|]$ are
\begin{eqnarray}\label{eigen_values_lambda_2q_process}
\nonumber \lambda_0&=&(\chi_0^2+\chi_1^2+\chi_2^2+\chi_3^2)^2,\\
\nonumber \lambda_1&=&4(\chi_1\chi_2+\chi_3\chi_0)^2,\\
\nonumber \lambda_2&=&4(\chi_1\chi_3+\chi_2\chi_0)^2,\\
\lambda_3&=&4(\chi_1\chi_0+\chi_2\chi_3)^2.
\end{eqnarray}
$\lambda_0$ is the maximal eigenvalue of $\Lambda$. Substituting the
values of $R_i$ from Eq. (\ref{Radii chi eigenvalues}) results with
the output state concurrence
\begin{equation}\label{two_qubit_dynamics}
C[\mathcal{E}(|\psi_B\rangle\langle\psi_B|)]=\textrm{max}\{(R_1^2+
R_2^2+R_3^2-1)/2,0\}.
\end{equation}
Notice the similarity between our result of Eq.
(\ref{two_qubit_dynamics}) and the former result of Eq.
(\ref{one_qubit_dynamics}). Accordingly, the entanglement breaking
condition in the last case is $R_1^2+R_2^2+R_3^2\leq1$. This
condition was reported in \cite{Filippov} as applicable to the case
of unital channels that are applied on both qubits of any entangled
state. Unlike Eq. (\ref{one_qubit_dynamics}), the entanglement
dynamics described by Eq. (\ref{two_qubit_dynamics}) does not apply
to \textit{any} unital channel and to \textit{any} maximally
entangled initial state. Numerical simulations suggest that in the
general case where the unital channel is not a Pauli channel, the
loss of entanglement is faster, and $(R_1^2+R_2^2+R_3^2-1)/2$ is
only an upper bound for the concurrence. This was only proved for
the entanglement breaking point, but not for channels that may leave
the state partially entangled \cite{Filippov}. Equation
(\ref{two_qubit_dynamics}) does hold in some cases, such as when the
noise is isotropic ($R_1=R_2=R_3$, as will be shown below) and the
initial state is any maximally entangled state, or for any unital
noise when the initial state is the singlet state $|\psi^-\rangle$.

\section{Experimental setup}
\begin{figure}[tbp]
\includegraphics[angle=0,width=86mm]{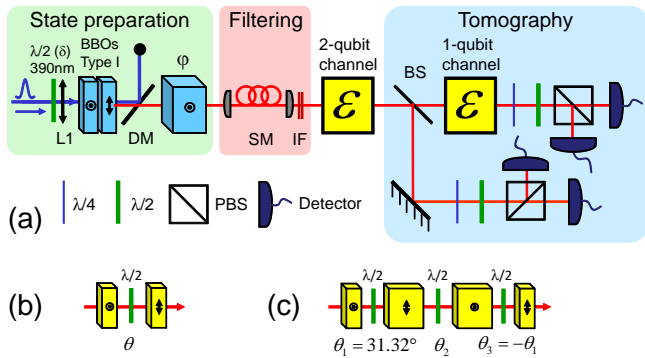}
\caption{\label{Fig1}(color online) The experimental setup. (a)
State generation and characterization units: photon pairs are
generated in the BBO crystals, which are located after a lens (L1)
and a half-wave plate (HWP, $\lambda/2$) in an angle of $\delta$.
The down-converted signal passes through a dichroic mirror (DM), a
birefringent compensating crystal $(\varphi)$, a single-mode fiber
(SM), and a 3\,nm interference bandpass filter (IF). In the state
characterization unit, the photons are split probabilistically by
a beam splitter (BS) to two ports. In each port, the photons pass
a sequence of a quarter-wave plate ($\lambda/4$), a HWP, and a
polarizing beam splitter (PBS) before being coupled into
single-photon detectors. (b) The two-field channel, composed of
two perpendicularly oriented identical 2\,mm thick calcite
crystals. (c) The three-field (isotropic) channel. This channel is
composed of four crystals and two fixed HWPs. The thickness of the
two outer (inner) crystals is 1\,mm (2\,mm).}
\end{figure}

In order to study entanglement dynamics experimentally, we generated
entangled pairs of photons, transmitted them through controllable
unital noisy channels, and measured the final concurrence of the
output states. Entanglement was formed between the polarization
degrees of freedom (DOFs), where the horizontal $|h\rangle$ and the
vertical $|v\rangle$ polarization modes define the computational
basis. The experimental setup is shown in Fig. \ref{Fig1}(a): Photon
pairs are collinearly generated by the process of spontaneous
parametric down conversion. Using a lens of 30\,cm focal length
(L1), a pulsed 390\,nm pump laser is focused onto two
perpendicularly oriented 1\,mm thick type-I
$\beta-\textrm{BaB}_{2}\textrm{O}_{4}$ (BBO) crystals. After the
crystals, the down-converted signal is separated from the pump beam
using a dichroic mirror (DM). A half-wave plate (HWP) at an angle of
$\delta$ is placed before the generating crystals in order to
control the relative pump power of each crystal. Thus, the generated
state is $\
|\psi\rangle=\cos(2\delta)|hh\rangle+\sin(2\delta)e^{i\varphi}|vv\rangle$.
The angle $\varphi$ is controlled by tilting another compensating
crystal which is placed after the generating crystals. Before
entering the quantum channel, the state is filtered spatially using
a single-mode fiber (SM), and spectrally by a 3\,nm interference
bandpass filter (IF). In the state characterization unit, the
photons are first split probabilistically by a beam splitter (BS).
Then, their post-selected two-port polarization state is measured by
a two-qubit quantum state tomography procedure \cite{Kwiat_Tomo}.
The required projections are achieved by wave-plates and polarizing
beam splitters (PBS) that are placed before the single-photon
detectors of each port.

Controlled quantum channels were implemented using a sequence of
fixed birefringent calcite crystals and HWPs
\cite{Kwiat_Dephasing,Shaham} (see Fig. \ref{Fig1}(b),(c)). Each
crystal entangles the polarization modes of a photon with its
internal temporal DOFs. Decoherence occurs when the photon detection
is insensitive to the temporal delays, practically averaging over
these DOFs. In order to apply a channel to one of the qubits we
placed it in one port after the BS. A two-qubit channel was realized
by placing the channel before the BS. For both channel types,
control over the noise probability $P$ was achieved by rotating the
corresponding HWPs to different angle settings \cite{Shaham}.

Two different unital channels were implemented. The first is the
two-field channel \cite{Shaham,Shaham_QPT}, (see Fig.
\ref{Fig1}(b)). It is described by random, but equally probable
$\sigma_1$ or $\sigma_2$ rotations of the initial state, with
overall noise probability of $P$
\begin{equation}\label{two_field_channel}
\mathcal{E}(\hat{\rho})=(1-P)\hat{\rho}+\frac{P}{2}\sigma_1\hat{\rho}\sigma_{1}+\frac{P}{2}\sigma_2\hat{\rho}\sigma_{2}\,.
\end{equation}
Substituting the process matrix eigenvalues
$\{1-P,\frac{P}{2},\frac{P}{2},0\}$ in Eq. (\ref{Radii chi
eigenvalues}), we find that the primary radii of the mapped
ellipsoid follows $\{R_1=R_2,R_3=2R_1-1\}$. The second channel is an
isotropic depolarization channel \cite{Shaham_iso_depo} (see Fig.
\ref{Fig1}(c)). It is described as
\begin{equation}\label{isotropic_channel}
\mathcal{E}(\hat{\rho})=(1-P)\hat{\rho}+\frac{P}{3}\sigma_1\hat{\rho}\sigma_{1}+\frac{P}{3}\sigma_2\hat{\rho}\sigma_{2}+\frac{P}{3}\sigma_3\hat{\rho}\sigma_{3}\,.
\end{equation}

\begin{figure}[tbp]
\includegraphics[angle=0,width=86mm]{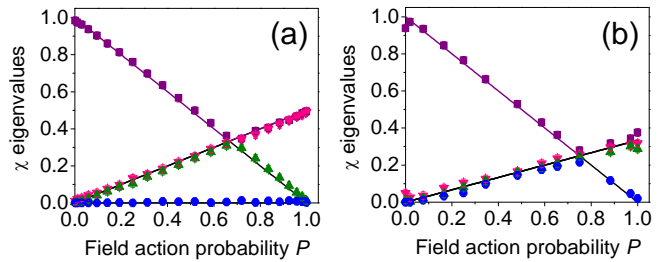}
\caption{\label{Fig2}(color online) Channel characterization.
Measured eigenvalues of the process matrices for the (a) two-field
and (b) the isotropic channels as a function of the noise
probability $P$, with their theoretical predictions (solid lines).}
\end{figure}

The eigenvalues of the single-qubit process matrix of these
channels, as obtained with a quantum process tomography procedure
are presented in Figs. \ref{Fig2}(a) and \ref{Fig2}(b). The noise
probability $P$ is controlled by the rotation of the corresponding
channel wave-plate. Surprisingly, for both channels
$P=\textrm{sin}^2(2\theta)$, where for the isotropic channel
$\theta=\theta_2$. Errors are calculated using a maximal likelihood
procedure and Monte Carlo simulations, assuming Poissonian noise
\cite{Kwiat_Tomo,Shaham_QPT}.

\section{Dynamics of maximally entangled states}
\begin{figure}[tbp]
\includegraphics[angle=0,width=70mm]{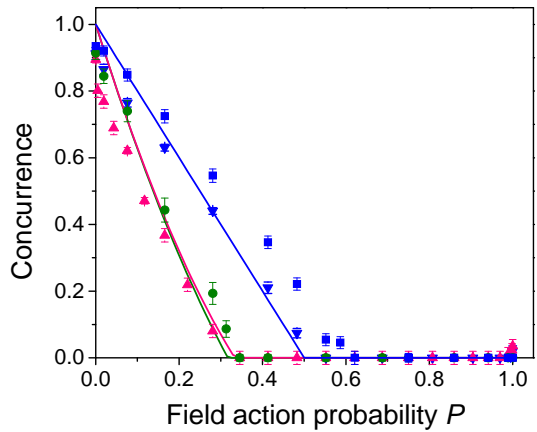}
\caption{\label{Fig3}(color online) Measured entanglement dynamics
of a $|\phi^+\rangle$ state. Four cases are presented: The two-field
channel, when applied to one photon (downward blue triangles) or to
both (upward pink triangles), and the isotropic channel, when
applied to one photon (blue squares) or to both (green circles).
Solid lines represent the theoretical predictions.}
\end{figure}

\begin{figure}[tbp]
\includegraphics[angle=0,width=86mm]{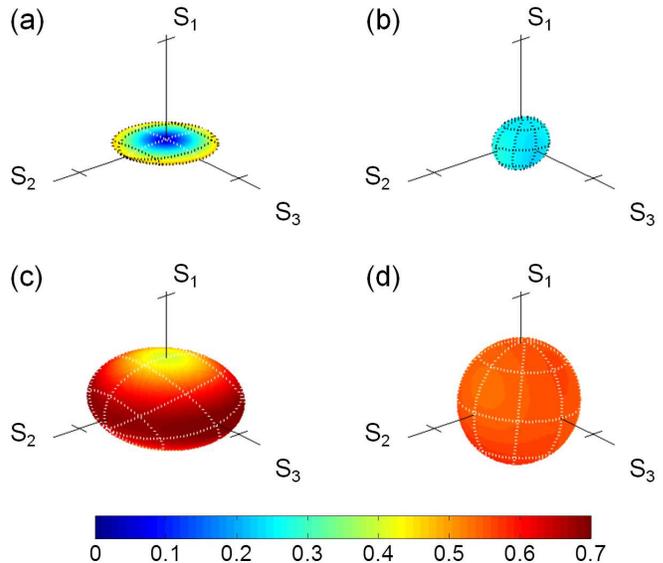}
\caption{\label{Fig4}(color online) Reconstruction of the measured
entanglement breaking processes in the Bloch sphere representation.
(a) and (b) correspond to the two-field and the isotropic channels
when applied to only one of the photons, respectively. (c) and (d)
represent the cases when the two-field and the isotropic channels
are applied to both photons, respectively. Ellipsoid surface colors
show the distance from the origin. Axis ticks are at values of 1. }
\end{figure}

Setting $\delta=22.5^\circ$ and $\varphi=0$, we generated
$|\phi^+\rangle$ Bell states with an initial concurrence of
$0.90\pm0.01$. Either one or both photons were transmitted through
the two types of channels, as described before. For each decoherence
setting, the output state concurrence is calculated from the
reconstructed density matrix. It is presented in Fig. \ref{Fig3} as
a function of the noise probability $P$, along with the theoretical
predictions. When either the two-field or the isotropic channels are
applied to one of the two qubits, the concurrence is degrading
similarly as a linear function of $P$. Entanglement breaking should
occur when $P=\frac{1}{2}$. Reconstructed processes of the measured
entanglement breaking points are presented in Fig. \ref{Fig4} using
the Bloch sphere representation: A two-field process of
$P=0.52\pm0.01$ is shown in Fig. \ref{Fig4}(a), and an isotropic
process of $P=0.59\pm0.01$ is shown in Fig. \ref{Fig4}(b). According
to theoretical calculations, if the channels are applied to both
qubits, the concurrence dynamics is quadratic with $P$. Here, for
the two processes, the dynamics is close but not identical.
Entanglement breaking for the two-field channel occurs when
$P=\frac{1}{3}$, and for the isotropic channel when
$P=\frac{3-\sqrt{3}}{4}\simeq0.317$, which corresponds to a mapped
sphere with a radius of $R=\sqrt{\frac{1}{3}}$. The corresponding
measured processes for the two-field channel $(P=0.31\pm0.01)$, and
for the isotropic channel $(P=0.35\pm0.01)$ are presented in Fig.
\ref{Fig4}(c) and Fig. \ref{Fig4}(d), respectively.

As predicted, when the channels are applied to both photons,
concurrence vanishes faster than when applied to one. The deviation
from theory is larger for the isotropic channel, where decoherence
is slower than expected. We explain this as a result of insufficient
temporal separation by the birefringent crystals
\cite{Kwiat_Dephasing,Shaham,Shaham_iso_depo}; The coherence time of
300\,fs is determined by the 3\,nm wide bandpass filters that were
used for spectral filtering. The two-field channel is using two
2\,mm wide calcite crystals. On the other hand, the four-crystal
configuration for the isotropic channel requires a different width
for two of the crystals. Thus, we also used in this case two 1\,mm
crystals that reduced and delayed the decoherence effects.

\section{Dynamics of initial partially entangled states}
In addition to studying the dynamics of initial maximally entangled
states, we investigated the evolution of partially entangled states
(PES). We now focus on the case of an isotropic channel which is
applied only to one of the qubits. Two different classes of initial
states were considered: pure PES and mixed PES. Different pure PES
are generated by adjusting the HWP angle $\delta$, so that the
crystals are not equally pumped. The initial state concurrence is
$C=\sin(4\delta)$. Specifically, we applied isotropic noise to
states with initial concurrence of $0.50\pm0.01$ and $0.16\pm0.01$.
When applied to pure PES, the concurrence dynamics should evolve
according to the factorization law, derived by Thomas \textit{et
al.} \cite{Konrad_2008}:
\begin{equation}\label{concurrence factorization}
C[(I\otimes\$)\hat{\rho}]=C[(I\otimes\$)|\phi^+\rangle\langle\phi^+|]C(\hat{\rho})\,.
\end{equation}
This relation states that the concurrence of any initial pure state
after the application of a channel \$ on one of the qubits can be
factorized into the initial state concurrence and the concurrence
that results when the same channel is applied to a pure Bell state.

\begin{figure}[tbp]
\includegraphics[angle=0,width=86mm]{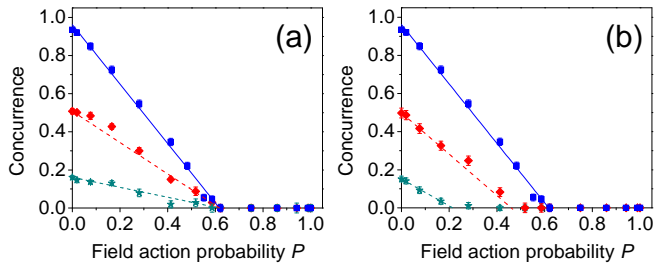}
\caption{\label{Fig5}(color online) Experimental results of the
entanglement dynamics when the isotropic channel is applied to one
qubit of initially (a) pure and (b) mixed PES.}
\end{figure}

The results for the initially pure PES are presented in Fig.
\ref{Fig5}(a), along with the corresponding dynamics of an initial
$|\phi^+\rangle$ state that also appears in Fig. \ref{Fig3}. The
solid line in Fig. \ref{Fig5}(a) represents a linear fit for the
measured dynamics of the $|\phi^+\rangle$ state. As was stated
before, theory predicts a linear dependency, and entanglement
breaking at $P=0.5$. Because of experimental errors, the concurrence
fit reaches zero only at $P=0.62\pm0.02$. From Eq. (\ref{concurrence
factorization}) it is clear that also the concurrence of the pure
PES should have a linear dependency on $P$. We draw the straight
dashed lines that connect the initial state concurrence at $P=0$ and
the experimental entanglement breaking point. As can be seen in Fig.
\ref{Fig5}(a), the dynamics of the two PES indeed follows the
corresponding linear predictions, reaching zero concurrence at the
same point.

We also studied initial PES that are partially mixed. The
entanglement evolution of such states was derived as an extension to
Eq. \ref{concurrence factorization} by Jim\'{e}nez-Far\'{i}as
\textit{et al.} \cite{Jimenez_2009}. The initial mixed PES
$\hat{\rho}$ is expressed in terms of an additional channel $\$'$
that is applied on one qubit of a pure two-qubit state
$\hat{\sigma}$ which is not necessarily maximally entangled:
$\hat{\rho}=(I\otimes\$')\hat{\sigma}$ \cite{Werner_book}. The
concurrence of $\hat{\rho}$ after a \$ channel is applied to one of
its qubits is
\begin{equation}\label{concurrence evolution_law_mixed_states}
C[(I\otimes\$)\hat{\rho}]=C[(I\otimes\$\$')|\phi^+\rangle\langle\phi^+|]C(\hat{\sigma})\,.
\end{equation}

Mixed PES were generated by introducing partial dephasing to an
initial $|\phi^+\rangle$ state as follows: the compensating
birefringent crystal was replaced with shorter crystals that did not
correct sufficiently the temporal walk-off between the horizontal
and the vertical amplitudes of the initial state, effectively
generating a dephasing noise \cite{Xu_2009,Kwiat_Dephasing}. in our
case, where $\$'$ is a dephasing channel, and $\$$ is an isotropic
channel, the concurrence also has a linear dependency on the
isotropic noise probability $P$. Compared to the former case of
initial pure states, decoherence occurs faster. As the initial state
is more dephased, it will lose its entanglement earlier.

The experimentally measured dynamics of two initially mixed PES with
initial concurrence of $0.50\pm0.03$ and $0.15\pm0.01$ is presented
in Fig. \ref{Fig5}(b), together with the $|\phi^+\rangle$ previous
results. The initial concurrence values are similar to those that
were studied in the pure PES case. The solid straight line is the
same fit to the $|\phi^+\rangle$ case as with the pure PES case. The
theoretical lines for the other two cases were calculated
numerically for the concurrence values of the initially mixed PES.
The presented dashed lines are corrected according to the
experimental deviation of the $|\phi^+\rangle$ case, i.e., their $P$
values are multiplied by the ratio $0.62/0.5$ between the measured
and predicted $P$ values for entanglement breaking.

The presented results demonstrate that the entanglement contained in
mixed PES is more fragile to noise compared to that of pure PES with
the same level of concurrence. As in the case of maximally entangled
states (Fig. \ref{Fig3}), most of the experimental deviation from
theory is explained by the length of the shortest crystals of the
isotropic channel. Additional deviation in the PES case results from
the overlap between the temporal modes of the initial dephasing
channel and those created by the isotropic channels. Nevertheless,
our results demonstrate the differences between the various cases
very clearly.

\section{Conclusions}
In this work, we studied the dynamics of entangled states when
subjected to unital noisy channels. We showed that concurrence, as
an entanglement measure, is linked in a simple way to the principal
radii of the ellipsoid that represent the noise map in the Bloch
sphere picture. Explicitly, when the channel is applied on one of
the qubits, the concurrence is linear with the sum of the absolute
values of the ellipsoid radii. We derived the concurrence for the
case when the same Pauli channel is applied to both qubits, and
found that it has a similar dependence, but with the sum of the
squares of the ellipsoid radii.

We realized two different channels using birefringent crystals.
These channels were applied to either one or to both photons of a
polarization entangled photon pair, in order to experimentally
demonstrate the entanglement dynamics of maximally entangled
polarization states. In the case of isotropic noise that is applied
to one of the photons, we also compared the entanglement degradation
of initially pure and mixed partially entangled states. For states
of similar initial concurrence, we have shown that dynamics depend
on the initial degree of mixedness.

Two interesting issues that we leave open are the generalization of
these results to non-unital channels and for systems of higher
dimensionality. One may speculate: suppose a known unital channel
operates on a maximally entangled three-qubit state. Can we quantify
the entanglement of the output state using an entanglement measure
which is proportional to the sum of the cubes of the ellipsoid radii
that describe this channel?

\section*{Acknowledgments}
We thank the Israeli Ministry of Science and Technology for
financial support.

\end{document}